\documentclass[twocolumn,preprintnumbers,secnumarabic,amsmath,amssymb,nofootinbib,floatfix]{revtex4}
\usepackage{varioref,exscale,latexsym,amsmath,amssymb}
\usepackage{graphicx}

\usepackage{graphicx}
\usepackage{dcolumn}
\usepackage{bm}

\def\beq{\begin{equation}}

\def\eeq{\end{equation}}

\def\beqa{\begin{eqnarray}}

\def\eeqa{\end{eqnarray}}

\begin{document}

\title{{\bf Limiting velocities as running parameters and superluminal neutrinos }}

\medskip\
\author{Mohamed M. Anber${}^{1}$}
\email[Email: ]{manber@physics.utoronto.ca}
\author{ John F. Donoghue${}^{2}$}
\email[Email: ]{donoghue@physics.umass.edu}
\affiliation{${}^{1}$Department of Physics,
University of Toronto\\
Toronto, ON, M5S1A7, Canada\\
${}^{2}$Department of Physics,
University of Massachusetts\\
Amherst, MA  01003, USA\\
}

\begin{abstract}
In the context of theories where particles can have different limiting velocities, we review the running of particle speeds towards a common limiting velocity at low energy. Motivated by the recent OPERA experimental results, we describe a model where the neutrinos would deviate from the common velocity by more than do other particles in the theory, because their running is slower due to weaker interactions.
\end{abstract}
\maketitle


\section{Introduction}
Recently, we \cite{Anber:2011xf} considered theories where the particles have different limiting velocities. Due to interactions, these velocities become running parameters and we showed that the limiting velocities are driven by the renormalization group (RG) equations towards a universal limiting velocity - the ``speed of light'' - at low energy. Because the running is produced by the self-interaction diagrams involving the other particles in the theory, we noted that the differences between the limiting speeds is greatest when the interactions are the weakest, and we pointed to gravitational waves as potentially the most sensitive test of this idea.

Recently the OPERA collaboration has presented evidence that neutrinos may have a different limiting velocity than the speed of light \cite{:2011zb}. Neutrinos are the second weakest interacting particles, after gravitons. While we are aware of potential problems with the experiment\cite{Contaldi:2011zm} and potential difficulties for the idea of a neutrino speed that is different from $c$ \cite{Cohen:2011hx, Cacciapaglia:2011ax, OPERA}, we would like to address here a mechanism for allowing the neutrinos to have a greater velocity difference from $c$ than the constraints on other particles would otherwise appear to allow.

The ideas of renormalization group evolution of the limiting speeds and the general approach towards a common speed at low energy are independent of the underlying mechanism, as long as one has an effective four-dimensional theory of particles with different speeds interacting with each other. However, our original motivation involved emergent fields. The Weinberg-Witten theorem is usually interpreted as telling us that non-Abelian gauge bosons and gravitons cannot be emergent fields arising from any underlying Lorentz invariant non-gauge theory. Such a Lorentz-violating underlying theory would generally involve wave equations which have different limiting velocities, so it would be a puzzle as to why Lorentz invariance appears as such a good symmetry at low energy. Our work using the renormalization group \cite{Anber:2011xf} is a partial answer to the issue of an emergent Lorentz symmetry.

In section 2 we briefly review the running of the limiting speeds. In section 3 we present a model in which the neutrino limiting speed differs from $c$ by a greater amount than does that of the charged leptons. A brief discussion completes this note.

\section{Limiting velocities as running parameters}

In a theory with Lorentz invariance, there is a common limiting speed $c$ and this speed can be used to define a natural set of units in which $c=1$. However if the theory does not have a full Lorentz invariance, the wave equations for the various fields can involve limiting velocities $c_i$ that are different from each other. Because these velocities are no longer universal, the limiting speeds cannot be universally removed from the theory and differences among them become important. Indeed, quantum corrections may modify the propagation of all the particles through the self-energy diagrams.

Self-energy diagrams generally involve more than one type of field. For example, in QED the photon self-energy involves fermions and the fermion self-energy involves a photon in addition to a fermion. Due to the lack of universal Lorentz invariance in such a theory, the time-like component of the propagator and the space-like component get renormalized differently. Because the limiting speed is related to the coefficients of energy and momentum variables, this means that the limiting speed itself gets renormalized, and the renormalization depends on the limiting velocity of both the fermion ($c_f$) and the gauge boson ($c_g$). Lorentz invariance would guarantee that there is no running of these parameters in the limit that $c_f=c_g$, so that we know that the running will be proportional to $c_f-c_g$ in the limit of small differences. The parameters do run when $c_f-c_g \neq 0$. To decide whether the speeds run towards each other, or away from each other, at low energy require
 s explicit calculations.

All of the cases that we treated in \cite{Anber:2011xf} resulted in the speeds approaching a common velocity at low energy. For example, the case of an interacting fermion and $U(1)$ gauge boson resulted in the beta functions
\begin{eqnarray}
\nonumber
\beta(c_g)&=&\frac{4e^2}{3\left(4\pi\right)^2}\frac{\left(c_g^2-c_f^2\right)}{c_fc_g}\,,\\
\beta(c_f)&=&\frac{8e^2}{3\left(4\pi\right)^2}\frac{\left(c_f-c_g\right)\left(4c_f^2+3c_fc_g+c_g^2\right)}{c_g\left(c_f+c_g\right)^2}\,.
\end{eqnarray}
These lead to $c_f=c_g$ as an infrared attractive fixed line.

The rate of approach towards the fixed point depends on the strength of the interaction and the number of fields involved. For a single field, the
approach is logarithmic, which raises the issue of whether approach is sufficient to satisfy experimental constraints. In \cite{Anber:2011xf} we provided a model with more fields, elaborated below, in which the running was much faster. The issue of whether the running would be logarithmic over large energy scales, or faster over a narrower energy range, is model dependent.

\section{The model}

One of the proposals to account for superluminal neutrinos is to use Coleman-Glashow idea \cite{Coleman:1997xq} that any particle will have its own mass as well as maximum attainable velocity. Although this scenario is in perfect agreement with the long baseline neutrino experiments with $v_{\nu}/c_\gamma \approx 1+10^{-5}$, where the best fit with the data leads to energy independent profile \cite{Cacciapaglia:2011ax}, it does not tell us why  there is a stringent bound on the difference between the maximum attainable velocity of charged leptons and speed of light \cite{Altschul}
\begin{eqnarray}
\left|1-\frac{c_e^2}{c_\gamma^2}\right|<10^{-14}\,.
\end{eqnarray}

Here we argue that if Lorentz symmetry is an emergent symmetry with a large number of fields, one can explain the hierarchy between the limiting speeds of electrons and neutrinos.

\begin{figure}[ht]
\leftline{
\includegraphics[width=.48\textwidth]{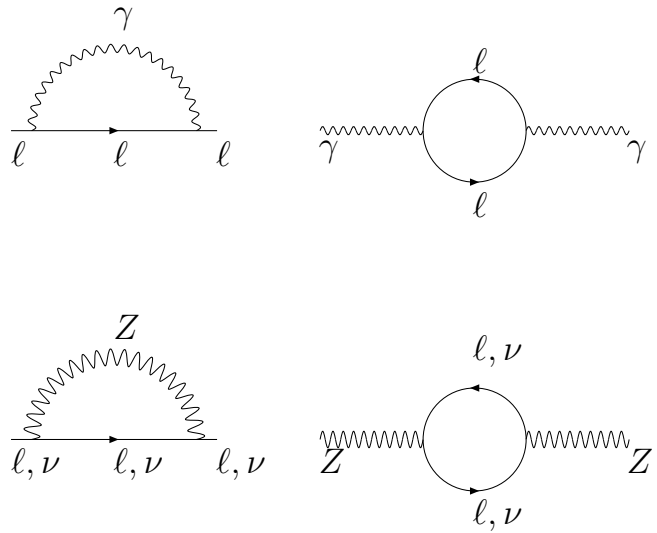}
}
\caption{Lower order quantum loops for $N_\gamma$ photons and $N_Z$ neutral Z's interacting with $N_\ell$ leptons.}
\label{quantum loops}
\end{figure}

One of the keys to this puzzle is to notice that neutrinos, unlike all other leptons, are not charged under $U_{em}(1)$. Now, let us introduce a large number $N_{\gamma}$ of hidden $U(1)$ photons as well as a large number $N_Z$ copies of neutral Zs (under the $U(1)$s) in addition to the Standard Model (SM) ones \cite{Dvali}. Moreover, we assume that there are a few leptons $N_{\ell\nu}$ (hidden and SM) that we assume for simplicity have the same origin, and hence all have the same initial speed of light $1+c_{f}^*$, with $|c_{f}^*|<<1$, at some UV emergent scale $\mu^*$. The charged leptons, denoted by $\ell$, have a common initial charge $Z^{*}_{\ell\ell,\gamma}$ under the different $U(1)$'s gauge sectors, while both charged and neutral leptons (the neutral ones are denoted by $\nu$) are charged initially with charge $Z^{*}_{\ell\nu,Z}$ under the $N_Z$ copies of $Z$s. The gauge bosons emerge with some initial speed $1+c_{g^*}$, with $|c_{g^*}|<<1$, that is different from the
  speed of fermions. At the UV scale the fermions and gauge bosons are taken to be massless and hence will participate in the RG running. As we run down our RG equations, most of the hidden particles become massive and decouple from the RG system. This situation is illustrated in FIG. \ref{quantum loops}. Here we do not assume any specific model for the nature of the coupling between the leptons and Zs. All we need to assume is the existence of a positive sign $\beta$ function for $\beta(Z_{\ell\nu,Z})$.  In order to produce the required hierarchy we take $N_\gamma>>N_Z>> N_{\ell}$.

Since all fermions share a common initial speed (also all gauge bosons have their common initial speed) and common initial coupling strengths, the evolution of the system can be modeled with only six parameters: $c_\gamma$, $c_Z$, $c_\ell$, and $c_\nu$ are respectively the photon, Z, charged lepton, and neutrino speeds, in addition to the coupling strengths $Z_{\ell\ell,\gamma}$ and $Z_{\ell\nu,Z}$.  Writing the speed of species $i$ as $c_i=1+\delta_i$, we find that the (RG) equations read
\begin{eqnarray}
\label{gamma}
\beta_{\delta_\gamma}&=&g_1N_\ell Z_{\ell\ell,\gamma}^{2}\left(\delta_\gamma-\delta_\ell\right)\\
\label{Z}
\beta_{\delta_Z}&=&g_2N_\ell Z_{\ell\nu,Z}^{2}\left(\delta_Z-\delta_\ell\right)+g_3N_\ell Z_{\ell\nu,Z}^{2}\left(\delta_Z-\delta_\nu\right)\\
\label{ell}
\beta_{\delta_\ell}&=&g_4N_\gamma Z_{\ell\ell,\gamma}^{2}\left(\delta_\ell-\delta_\gamma\right)+g_5 N_z Z_{\ell\nu,Z}^2\left(\delta_\ell-\delta_Z\right)\\
\label{nu}
\beta_{\delta_\nu}&=&g_6N_ZZ_{\ell\nu,Z}^2\left(\delta_\nu-\delta_Z \right)\\
\label{ellellgamma}
\beta_{Z_{\ell\ell,\gamma}}&=&g_7N_\ell Z_{\ell\ell,\gamma}^3\\
\label{ellnu}
\beta_{Z_{\ell\nu,Z}}&=&g_8N_\ell Z_{\ell\nu,Z}^3\,,
\end{eqnarray}
where $\beta_{\delta_\gamma}\equiv \left(4\pi \right)^2\mu d\delta_{\gamma}/d\mu$ atc., and $\{g_i\}$ are ${\cal O}(1)$ loop-numbers that can be calculated in any realistic setup.

As was shown in \cite{Anber:2011xf}, the positivity of $\{g_i\}$ suffice for the existence of a stable IR fixed line $c_\gamma=c_Z=c_\ell=c_\nu$. In the following we perform a series of approximations that enable us to show our main point. Namely, the hierarchy $N_\gamma>>N_Z>> N_{\ell}$ leads to a hierarchal structure of the different speeds with respect to the speed of light. Some of these approximations will turn out to be crude when compared to numerical integration of the whole system. However, this procedure is sufficient to illustrate our point. Subtracting Eqs. \ref{gamma} and \ref{ell}, \ref{Z} and  \ref{ell}, and \ref{Z} and \ref{nu}, and using $N_\gamma>>N_Z>> N_{\ell}$, therefore neglecting the less dominating terms, we obtain
\begin{eqnarray}
\label{ellMgamma}
\beta\left(\delta_\ell-\delta_\gamma\right)&\approx&g_4N_\gamma Z_{\ell\ell,\gamma}^2\left(\delta_\ell-\delta_\gamma\right)\,,\\
\nonumber
\beta\left(\delta_\ell-\delta_Z\right)&\approx&g_5N_ZZ_{\ell\nu,Z}^2\left(\delta_\ell-\delta_Z\right)\\
&&+g_4N_\gamma Z_{\ell\ell,\gamma}^2\left(\delta_\ell-\delta_\gamma\right)\,,\\
\label{nuMZ}
\beta\left(\delta_\nu-\delta_Z\right)&\approx&g_6N_ZZ_{\ell\nu,Z}^2\left(\delta_\nu-\delta_Z\right)\,.
\end{eqnarray}
Finally, we compute $\beta\left(\delta_\nu-\delta_\gamma\right)$ by noticing
\begin{eqnarray}
\nonumber
\beta \left(\delta_\nu-\delta_\gamma  \right)&=&\beta \left(\delta_\nu-\delta_Z  \right)+\beta \left(\delta_Z-\delta_\ell  \right)\\
\nonumber
&&+\beta \left(\delta_\ell-\delta_\gamma  \right)\\
\nonumber
&\approx&g_6N_ZZ_{\ell\nu,Z}^2\left(\delta_\nu-\delta_Z\right)\\
\label{additional equation}
&&-g_5N_ZZ_{\ell\nu,Z}^2\left(\delta_\ell-\delta_Z\right)\,.
\end{eqnarray}
Eqs. \ref{ellMgamma} to \ref{additional equation} as well as \ref{ellellgamma} and \ref{ellnu} can be integrated directly to obtain
\begin{eqnarray}
\nonumber
\eta_\ell\equiv\left|\frac{\delta_\ell-\delta_\gamma}{\delta_\ell^*-\delta_\gamma^*}\right|&\sim&\left(\frac{Z_{\ell\ell,\gamma}^2(\mu)}{Z_{\ell\ell,\gamma}^{*2}}\right)^{(N_\gamma/N_\ell)\times{\cal O}(1)}+\mbox{\scriptsize corrections}\,,\\
\nonumber
\eta_\nu\equiv\left|\frac{\delta_\nu-\delta_\gamma}{\delta_\nu^*-\delta_\gamma^*}\right|&\sim&\left(\frac{Z_{\ell\nu,Z}^2(\mu)}{Z_{\ell\nu,Z}^{*2}}\right)^{(N_Z/N_\ell)\times{\cal O}(1)}+\mbox{\scriptsize corrections}\,,\\
\label{crude approx}
\end{eqnarray}
where, for example,
\begin{eqnarray}
\frac{Z_{\ell\ell,\gamma}^{2}(\mu)}{Z_{\ell\ell,\gamma}^{*2}}=\frac{1}{1+\frac{2g_7N_\ell Z_{\ell\ell,\gamma}^{*2}}{(4\pi)^2}\log(\mu^*/\mu)}\,.
\end{eqnarray}
It turns out that the corrections in the above expression are important, despite being subleading, within the numerical simulations. However, this shows that charged leptons, that see $N_\gamma$, have more enhanced running than neutrinos, which only see $N_z<<N_\gamma$.

\begin{figure}[ht]
\leftline{
\includegraphics[width=.48\textwidth]{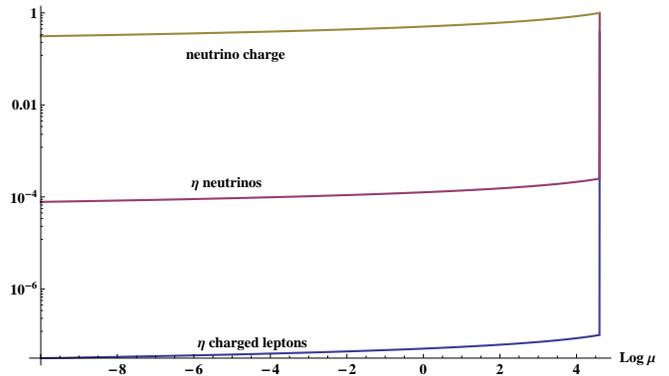}
}
\caption{Numerical simulation of the running of the charge $Z_{\ell\nu,Z}$, and ratios $\eta_\ell$ and $\eta_\nu$. We use the initial conditions $c_\nu^*=c_\ell^*=2$, $c_\gamma=c_Z=1$ and $Z_{\ell\nu,Z}^*=Z_{\ell\ell,\gamma}^*=1$ at $\mu^*=100$. We take $N_\gamma=5\times10^{8}$, $N_Z=2\times 10^5$, $N_\ell=50$, and $\{g_i\}=1$. We see that the running of $Z_{\ell\nu,Z}$ is logarithmic, while the running of $\eta_\ell$ and $\eta_\nu$ is power-law. Very small values of $\eta$ are achieved in a very short interval of running. Moreover, we see the hierarchical structure of the final speeds: $\eta_\ell<<\eta_\nu$.}
\label{RG flow of speeds and coupling}
\end{figure}

In FIG. \ref{RG flow of speeds and coupling} we demonstrate a numerical simulation of the RG flow of the system of Eqs. \ref{gamma} to \ref{ellnu}. One can see that by taking $N_\gamma>>N_Z>>N_\ell$ we can achieve $\eta_\ell<<\eta_\nu$.  Because the RG flow  of $Z_{\ell\nu,Z}$ is driven only by having a small number of fermionic species, the coupling constant runs only logarithmically. However, the huge number of gauge fields that participate in the RG equation of speeds for the charged particles force the running of $\eta$ to be extremely fast.

\begin{figure}[ht]
\leftline{
\includegraphics[width=.48\textwidth]{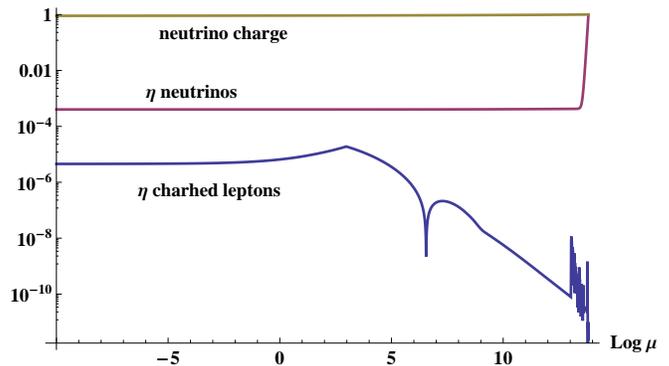}
}
\caption{Numerical simulation of the running of the charge $Z_{\ell\nu,Z}$, and ratios $\eta_\ell$ and $\eta_\nu$ using the power law dependence in Eq. \ref{power law running formulas}. We take the IR mass values $M_{\ell}=M_{\nu}=M_\gamma=1$, and we use $\Gamma_\gamma=0.02$, $\Gamma_{Z}=0.6$, $\Gamma_\ell=.5$, and $\alpha_\gamma=2.1$, $\alpha_Z=0.65$, $\alpha_\ell=0.1$. We use the initial conditions $c_\nu^*=c_\ell^*=2$, $c_\gamma^*=c_Z^*=1$ and $Z_{\ell\nu,Z}^*=Z_{\ell\ell,\gamma}^*=1$ at $\mu^*=10^6$, and we set $\{g_i\}=1$. We note that the value of $\eta_\ell$ is saturated by the error tolerance of the code. More powerful computations may give smaller values.}
\label{RG flow of speeds and coupling using power law formula}
\end{figure}

Although in the analysis conducted above we considered a constant number of hidden particles, this number may depend on the energy scale. We can model this dependence as a power law
\begin{eqnarray}
\label{power law running formulas}
N_{i}(\mu)&=&\Gamma_{i}\left(\frac{\mu}{M_{i}}\right)^{\alpha_{i}}\,,
\end{eqnarray}
where $\Gamma_{i}$ and $\alpha_{i}$ are positive constants, and $M_{i}$ is an IR mass scale. This exact behavior is also exhibited in models of large extra-dimensions where the Kaluza-Klein (KK) modes (from the 4D point of view) obey a power law as in Eq. \ref{power law running formulas} \cite{Masip:2000xy}.
\footnote{ It is also interesting to note the equivalence between theories with extra dimensions and strongly coupled QFT via the AdS/CFT correspondence.}
 In this context, $M_i$ is the lowest KK mode $M_i\sim 1/L$, where $L$ is the size of the extra dimension, and $\alpha=d$ is the number of extra dimensions. In FIG. \ref{RG flow of speeds and coupling using power law formula} we show the simulation of the RG system using the power-law formula in Eq. \ref{power law running formulas}.

To understand the choice of parameters used in the simulation in FIG. \ref{RG flow of speeds and coupling using power law formula}, it is instructive to calculate the total number of fermions and gauge fields as seen in the UV. Using Eq.  \ref{power law running formulas} and the numerical coefficients given in FIG. \ref{RG flow of speeds and coupling using power law formula} we find $N_\gamma\sim 10^{11}$,  $N_Z\sim 10^4$ and $N_\ell\approx 1$, which explains the hierarchical structure of the final speeds.

Now we comment on the validity of the RG equations in a perturbative treatment of QFT. It is inevitable that one must make sure  higher loop corrections are suppressed, otherwise our perturbative treatment (i.e. the loop-expansion) breaks down. As was stressed in \cite{Anber:2011xf}, the condition for the validity of perturbation theory puts a severe constraint on the number of species. This constraint may be relaxed if we notice that the condition for validity of perturbation theory is important mostly at the initial running, when  the charges $Z_{\ell\ell,\gamma}$ and $Z_{\ell\nu,Z}$ are relatively large. However, once the charges run down the low-energy scale, for example by means of non-perturbative RG treatment, the condition for validity of perturbation theory may be satisfied. Thus, as long as the non-perturbative result does not change the overall trend dramatically, fast running would be expected to still be obtained even if the perturbative analysis is not reliable
 in detail.

\section{Discussion}

If correct, the OPERA result would obviously have a dramatic effect on the
Lorentz symmetry of the fundamental interactions. However, it also raises two
subsequent problems. One is that the velocity difference of neutrinos
would have to be much greater than that allowed for many other particles. A second
problem is that the constraints of SN1987a force the velocity difference to be energy
dependent\cite{Cacciapaglia:2011ax}.

We have discussed a model in which the neutrino limiting velocities differ from $c$ by
exponentially larger factors than do those of the charged leptons. This is due to a
weaker running for neutrinos, caused by weaker interactions. The result of the evolution
is an energy independent speed in the
infrared. This model can therefore address one of the problems raised by OPERA, but does
not solve the issue of energy dependence. We present it in this note in the hopes that
the mechanism that we address could eventually find its way into a complete solution to
both issues, in the event that the OPERA result is ultimately confirmed.

\section*{Acknowledgements}

The work of M.A. is supported  by NSERC Discovery Grant of Canada. The work of J.D. has been supported in part by the NSF grant PHY - 0855119, and in part by the Foundational Questions Institute.

\end{document}